\input harvmac
\input epsf.tex
\input amssym.tex

\def\PP{{\Bbb{P}}}

\def\figin{\epsfcheck\figin}\def\figins{\epsfcheck\figins}
\def\epsfcheck{\ifx\epsfbox\UnDeFiNeD
\message{(NO epsf.tex, FIGURES WILL BE IGNORED)}
\gdef\figin##1{\vskip0in}\gdef\figins##1{\hskip.0in}
\else\message{(FIGURES WILL BE INCLUDED)}%
\gdef\figin##1{##1}\gdef\figins##1{##1}\fi}
\def\DefWarn#1{}
\def\figinsert{\goodbreak\midinsert}
\def\ifig#1#2#3{\DefWarn#1\xdef#1{fig.~\the\figno}
\writedef{#1\leftbracket fig.\noexpand~\the\figno}%
\figinsert\figin{\centerline{#3}}\medskip\centerline{\vbox{\baselineskip12pt
\advance\hsize by -1truein\noindent\footnotefont{\bf
Fig.~\the\figno:} #2}}
\bigskip\endinsert\global\advance\figno by1}


\overfullrule=0pt

\lref\others{
  M.~Sakaguchi and K.~Yoshida,
  arXiv:0805.2661 [hep-th].
    W.~D.~Goldberger,
  arXiv: 0806.2867 [hep-th].
  J.~L.~B.~Barbon and C.~A.~Fuertes,
  JHEP {\bf 0809}, 030 (2008)
  [arXiv:0806.3244 [hep-th]].
  M.~Sakaguchi and K.~Yoshida,
  JHEP {\bf 0808}, 049 (2008)
  [arXiv:0806.3612 [hep-th]].
  W.~Y.~Wen,
  arXiv:0807.0633 [hep-th].
  Y.~Nakayama,
  arXiv:0807.3344 [hep-th].
 J.~W.~Chen and W.~Y.~Wen,
  arXiv:0808.0399 [hep-th].
  D.~Minic and M.~Pleimling,
  arXiv:0807.3665 [cond-mat.stat-mech].
  S.~Kachru, X.~Liu and M.~Mulligan,
  arXiv:0808.1725 [hep-th].
  S.~S.~Pal,
  arXiv:0808.3042 [hep-th];
  arXiv:0808.3232 [hep-th];
  arXiv:0809.1756 [hep-th].
  P.~Kovtun and D.~Nickel,
  arXiv:0809.2020 [hep-th].
  C.~Duval, M.~Hassaine and P.~A.~Horvathy,
  arXiv:0809.3128 [hep-th].
  S.~S.~Lee,
  arXiv:0809.3402 [hep-th].
  D.~Yamada,
  arXiv:0809.4928 [hep-th].
  F.~L.~Lin and S.~Y.~Wu,
  arXiv:0810.0227 [hep-th].
  S.~A.~Hartnoll and K.~Yoshida,
  arXiv:0810.0298 [hep-th].
    M.~Schvellinger,
  arXiv:0810.3011 [hep-th].

}

\lref\SonYE{
  D.~T.~Son,
  Phys.\ Rev.\  D {\bf 78}, 046003 (2008)
  [arXiv:0804.3972 [hep-th]].
}

\lref\HorowitzAY{
  G.~T.~Horowitz, J.~M.~Maldacena and A.~Strominger,
  Phys.\ Lett.\  B {\bf 383}, 151 (1996)
  [arXiv:hep-th/9603109].
}

\lref\BalasubramanianDM{
  K.~Balasubramanian and J.~McGreevy,
  Phys.\ Rev.\ Lett.\  {\bf 101}, 061601 (2008)
  [arXiv: 0804.4053 [hep-th]].
}

\lref\WittenZW{
  E.~Witten,
  Adv.\ Theor.\ Math.\ Phys.\  {\bf 2}, 505 (1998)
  [arXiv:hep-th/9803131].
}

\lref\HerzogWG{
  C.~P.~Herzog, M.~Rangamani and S.~F.~Ross,
  arXiv:0807.1099 [hep-th].
}
\lref\ItzhakiDD{
  N.~Itzhaki, J.~M.~Maldacena, J.~Sonnenschein and S.~Yankielowicz,
  Phys.\ Rev.\  D {\bf 58}, 046004 (1998)
  [arXiv:hep-th/9802042].
}

\lref\AdamsWT{
  A.~Adams, K.~Balasubramanian and J.~McGreevy,
  arXiv:0807.1111 [hep-th].
}

\lref\AlishahihaRU{
  M.~Alishahiha and O.~J.~Ganor,
  JHEP {\bf 0303}, 006 (2003)
  [arXiv:hep-th/0301080].
}

\lref\GimonXK{
  E.~G.~Gimon, A.~Hashimoto, V.~E.~Hubeny, O.~Lunin and M.~Rangamani,
  JHEP {\bf 0308}, 035 (2003)
  [arXiv:hep-th/0306131].
}

\lref\MaldacenaWH{
  J.~Maldacena, D.~Martelli and Y.~Tachikawa,
  arXiv:0807.1100 [hep-th].
}

\lref\AharonyTI{
  O.~Aharony, S.~S.~Gubser, J.~M.~Maldacena, H.~Ooguri and Y.~Oz,
  Phys.\ Rept.\  {\bf 323}, 183 (2000)
  [arXiv:hep-th/9905111].
}

\lref\AlishahihaCI{
  M.~Alishahiha, Y.~Oz and M.~M.~Sheikh-Jabbari,
  JHEP {\bf 9911}, 007 (1999)
  [arXiv:hep-th/9909215].
}

\lref\HorowitzWT{
  G.~T.~Horowitz and D.~L.~Welch,
  Phys.\ Rev.\  D {\bf 49}, 590 (1994)
  [arXiv:hep-th/9308077].
}


\Title{\vbox{\vskip 0.5cm
}}
 {\vbox{{}\medskip\centerline{Non-relativistic Branes} }}
\smallskip
\centerline{Luca Mazzucato${}^{1,2}$, Yaron Oz${}^1$ and  Stefan Theisen$^3$}
\smallskip
\bigskip
\centerline{\it ${}^1$Raymond and Beverly Sackler  School of Physics and Astronomy}
\smallskip
 \centerline{\it Tel-Aviv University, Ramat-Aviv 69978, Israel}
\medskip
\centerline{\it ${}^2$Simons Center for Geometry and Physics}
\smallskip
 \centerline{\it Stony Brook University, Stony Brook, NY 11794-3840, USA}
\medskip
\centerline{\it ${}^3$Max-Planck-Institut f\"ur Gravitationsphysik, Albert-Einstein-Institut}
\smallskip
\centerline{\it 14476 Golm, Germany}
\medskip

\bigskip
\vskip 0.5cm

\centerline{}
{\noindent We construct gravitational non-relativistic  brane solutions
of type IIA/IIB string theories and M-theory and their near-horizon
geometries. The
non-relativistic M2 and M5-brane metrics have Schr\"{o}dinger symmetries
with dynamical exponent $z=3/2$ and $z=3$, respectively, whereas the
known D3-brane solution has $z=2$.
The non-relativistic D-brane, NS5-brane, F-string and KK monopole metrics
have asymptotically space and time translations, space rotations,
Galilean symmetries and
a particle number symmetry. We construct two different gravitational
backgrounds of a non-relativistic D1-D5 system, both having asymptotically
Schr\"{o}dinger symmetry with exponent $z=2$. We study the properties
of the solutions and their phase diagram.}
\vskip 0.5cm

\noindent

 \vskip 0.5cm

\Date{October 2008}


\newsec{Introduction}

The AdS/CFT correspondence relates conformal field theories in $d$-dimensional flat space-times
to gravitational theories (superstring/M-theory) in asymptotically $AdS_{d+1}$ curved spaces (for a review see \AharonyTI).
The gravitational description is weakly coupled when the conformal field theory is strongly coupled.
The AdS/CFT correspondence has been generalized in various ways in order to relate non-conformal field theories
to gravity. Two such generalizations have been realized by relating the worldvolume field theories of type II Dp-branes ($p\neq 3$) and
NS5-branes to their gravitational description.

Recently, several attempts have been made to generalize the AdS/CFT correspondence to strongly coupled non-relativistic conformal systems \SonYE\BalasubramanianDM.\foot{Other recent discussions include \others.}
The symmetry group of such systems is the non-relativistic conformal group (the Schr\"{o}dinger group) and it should be realized as the isometry
group of the corresponding curved metrics.  The Schr\"{o}dinger group consists of translations in time and space, space rotations,
Galilean boosts, dilatation and particle number symmetry.
Such a solution of non-relativistic D3-branes has been obtained by applying a Null Melvin Twist to the relativistic D3-brane
solution \AdamsWT\MaldacenaWH\HerzogWG, and by a Penrose limit or a TsT transformation in \MaldacenaWH. In this case the metric has  a special conformal symmetry in addition to the Schr\"{o}dinger symmetry.
This procedure has also been used to construct non-extremal non-relativistic D3-brane solution.

It is natural to ask whether one can construct non-relativistic branes of different dimensionality, i.e. Dp-branes for general $p$, NS-branes,
M-branes and their intersections, and thus to  generalize further the proposal for a duality between non-relativistic systems and gravity.
The aim of this paper is to construct such solutions and discuss their properties.
In the various sections we analyze in detail the gravitational backgrounds at zero and finite temperature.
Let us briefly present in the following  some of the results at zero temperature.
In the table we summarize the various near-horizon metrics and the sections of the paper where we discuss them.

$$
\eqalign{
\matrix{&{\rm Metric}& {\rm Section}\cr&&\cr
&ds_{Dp}^2=\left({\rho_p\over r}\right)^{7-p\over2}\left[-{2\Delta^2\over r^2}dt^2+2dtd\xi+(dx^i)^2\right]
 +\left({\rho_p\over r}\right)^{p+1\over2}\left[dr^2+r^2d\Omega^2\right], & 2.1;\, 3 \cr&&\cr
 &ds_{M2}^2 = \left({\rho_{M2}\over r}\right)^2 \left[-{2\Delta^2 \over r}dt^2 + 2 dt d\xi + (dx^i)^2 + dr^2 \right] + \rho_{M2}^2 d\Omega_7^2, & 4.1\cr&&\cr
&ds_{M5}^2 = \left({\rho_{M5}\over r}\right)^2 \left[-{2\Delta^2 \over r^{4}}dt^2 + 2 dt d\xi + (dx^i)^2 + dr^2 \right] + \rho_{M5}^2 d\Omega_4^2, & 4.2\cr&&\cr
 &ds^2_{NS5A}= -{2\Delta^2\over r^2 } dt^2+2dtd\xi+(dx_i)^2+{\rho_{NS5}\over r}\left(dr^2+r^2d\Omega_3^2\right), &4.4\cr&&\cr
&ds_{NS5B}^2=-{2\Delta^2\over r^2}dt^2+2dtd\xi+(dx^i)^2+\left({\rho_{NS5}\over r}\right)^2\left(dr^2+r^2 d \Omega_3^2 \right), &2.2\cr&&\cr
&ds^2_{F1}=
\left({\rho_{F1}\over r}\right)^6\left[-{2\Delta^2\over r^2}dt^2+2dtd\xi\right]+\left({\rho_{F1}\over r}\right)^4\left(dr^2+r^2 d \Omega_7^2 \right), &2.2; \, 4.4\cr&&\cr
&ds^2_{KK}=-{2\Delta^4\over  r^4}dt^2+2dtd\xi+(dx^i)^2+\left({\rho_{KK}\over  r}\right)^4ds^2_{ALE}, &4.3\cr&&\cr
 &ds^2_{D1D5}={\rho_1\rho_5\over r^2}\left[-\left(
{2\Delta^2 \over r^2}\right)dt^2+2dtd\xi+dr^2\right]+\rho_1\rho_5d\Omega_3^2+{\rho_1\over \rho_5}ds^2_{M_4}, & 5.1 \cr&&\cr
&ds^2_{F1NS5}={\rho_5^2\over r^2}\left[-\left(
{2\tilde\Delta^2 \over r^2}\right)dt^2+2dtd\xi+dr^2\right]+\rho_5^2d\Omega_{3}^2+ds^2_{M^4} , & 5.2
}}$$

Consider the non-relativistic Dp-branes, where the Dp-brane charge $Q_p \sim \rho_p^{7-p}$.
 The parameter $\Delta$ can be eliminated by a redefinition of the coordinates $t$ and $\xi$.
 Note, however, that this cannot be done for the non-extremal Dp-brane solutions and $\Delta$
 should be a physical parameter of the dual non-relativistic field theory related to a chemical potential \AdamsWT.
As for the relativistic Dp-branes, the dilaton is not constant when $p\neq 3$. There are two non-zero background fields: a RR $(p+1)$-form potential that takes the same form as in the relativistic case, and a NSNS 2-form B-field.

The metrics have along the worldvolume directions
space and time translations, space rotations and  Galilean boost symmetries
 $$x^i \rightarrow x^i -v^i t\ ,~~~~~~~\xi \rightarrow \xi + v^ix^i - {v^2 \over 2}t \ ,
 $$
in addition to a particle number symmetry (translation in $\xi$) and  an $SO(9-p)$ rotational symmetry in
 the $(9-p)$ transverse directions.
 The dual non-relativistic field theories live in the coordinates $(t,x^i)$, for $i=1,\dots,p-1$. The momentum in the $\xi$ direction is
interpreted as the particle number and the coordinate $r$ as the inverse RG length scale.
Note, however, that there is still an open issue of how to define properly the boundary where the non-relativistic field theory lives, since
the leading term of the metric in the worldvolume directions is one-dimensional as $r\rightarrow 0$.

When $p=3$, the metric  has two additional symmetries: a dilatation
 $$(t,\xi,x^i,r) \rightarrow (\lambda^2 t,\xi,\lambda x^i, \lambda r) \ ,
 $$
and a special conformal transformation
 $$(t,\xi,x^i,r) \rightarrow \left(t(1 -\lambda t),\xi + {\lambda \over 2}(x^2 + r^2),x^i(1 -\lambda t), r (1 -\lambda t)\right) \ .
 $$
There is only one (temporal) special conformal symmetry of the non-relativistic D3-branes metric, while there are four
special conformal symmetries in the relativistic case.

The non-relativistic M2 and M5-brane metrics have Schr\"{o}dinger symmetries that include also a dilatation. Note that
their metrics take the form
$$
 ds^2 = \left({\rho_A\over r}\right)^2 \left[-{2\Delta^2 \over r^{2\nu}}dt^2 + 2 dt d\xi + (dx^i)^2 + dr^2 \right] + \rho_A^2 d\Omega^2 \ .
 $$
$\nu$ is related to the dynamical critical exponent $z$ via $z=\nu+1$, where $z$ parametrizes the
anisotropy in the scaling of time and space coordinates: $t\to\lambda^z t,\, x\to \lambda x$.
In addition
there is a non-zero 3-form potential. Unlike the D3-brane case, special conformal
transformation is not a symmetry of the M2 and M5 branes metric.

The non-relativistic type IIB NS5-brane background has a linear dilaton,
a RR 2-form potential and a NSNS 2-form B-field.
It is obtained by applying an S-duality to the non-relativistic D5-brane background and has the same symmetries of the latter but it describes a coupled D1-NS5 system.
The non-relativistic type IIA NS5-brane background can be obtained by taking a type IIA limit of the non-relativistic M5-brane background.
The solution contains D2 branes stretched along a worldvolume direction and a transverse direction, thus describing a coupled D2-NS5 system.
The non-relativistic type II NS5-brane background provides a dual description of a non-relativistic Little String Theory.

The non-relativistic type IIB fundamental string background has a non-zero dilaton,
a RR 2-form potential and a NSNS 2-form B-field.
It is obtained by applying an S-duality to the D1-brane background and has the same symmetries. It describes a coupled D1-F1 system.
The non-relativistic type IIA fundamental string background is obtained by a reduction of the eleven-dimensional M2-brane background
to ten dimensions and has a non-zero dilaton,
a RR 3-form potential and a NSNS 2-form B-field, describing a coupled F1-D2 system.

Unlike the relativistic case, the non-relativistic D3-brane of type IIB is not invariant under S-duality, since the NSNS B-field is mapped
to a RR 2-form vector potential $C_2$. This implies that unlike ${\cal N}=4$ Super Yang-Mills, here the dual non-relativistic conformal field theory does not possess
an S-duality symmetry. Moreover, while the original background has a NSNS flux, the S-dual background is supported solely by RR fluxes.

Another eleven-dimensional non-relativistic background in the table is the Kaluza-Klein monopole obtained by lifting the $N$ D6-branes background
to eleven dimensions, where $ds_{ALE}^2$ is the metric on an $ALE_{N-1}$ space
and in addition there is a non-zero 3-form potential.

As for the non-relativistic D3-brane, the non-relativistic D1-D5 branes metric has a special conformal symmetry in addition to the Schr\"{o}dinger symmetry.
In the table
$M_4$ refers to the four-torus $T^4$ or to $K3$. In both cases, we can perform the shift along an isometry of the three-sphere and obtain a solution with a constant dilaton, NSNS B-field and RR 3-form field strength. If we pick the four-torus as a compact manifold, however, we can perform the shift along the torus isometry and obtain a different solution, with the same metric  but with an additional RR five form flux turned on.

The paper is organized as follows.
In section 2 we will consider the type IIB non-relativistic branes.
The non-relativistic Dp-brane backgrounds are obtained from the relativistic Dp-brane solutions
via a TsT transformation as in \MaldacenaWH,
or by using the Null Melvin Twist procedure as in \HerzogWG\AdamsWT, following \AlishahihaRU\GimonXK.
In the case of the type IIB Dp-branes it works as in the original example of \HerzogWG\AdamsWT\MaldacenaWH.
We verified that both methods generate the same non-relativistic Dp-brane backgrounds.
We use S-duality in order to obtain the non-relativistic type IIB NS5-brane and fundamental string backgrounds.
In section 3 we will consider the type IIA non-relativistic branes.
We will first perform a Null Melvin Twist of the IIA black branes and obtain non-relativistic metrics whose noncompact part depends explicitly on the coordinates of the sphere. We will then modify the solution to obtain an analogous solution to the IIB case, where the noncompact part of the metric is independent of the compact directions.
In section 4 we will consider the non-relativistic M-branes and KK monopole, which we will obtain by lifting the type IIA
non-relativistic D2, D4 and D6-branes to eleven dimensions. By reducing to ten dimensions the M2 and M5-branes
we obtain the IIA F1 and NS5-brane solutions.
In section 5 we will consider the D1-D5 system and present the two different solutions obtained by performing the TsT shift either along the three-sphere or along the compact torus. In section 6 we will study the phase diagrams of the various solutions. Finally, in the Appendix we collect some useful geometrical results.

\newsec{Non-relativistic Type IIB branes}

The relativistic non-extremal Dp-brane backgrounds in ten dimensions (IIA for p even, IIB for p odd) read (in the string frame and $l_s=1$)
 \eqn\solu{\eqalign{
 ds^2=&{1\over h}\left(-fd\tau^2+dy^2+(dx^i)^2\right)+h\left(f^{-1}d\rho^2+\rho^2d\Omega_{8-p}^2\right) \ ,\cr
 \star F_{p+2}=&{\rm Vol}(S^{8-p})Q_p \ ,\qquad e^{\Phi - \Phi_{\infty}}=h^{3-p\over2} ,\cr
}}
where ${\rm Vol}(S^{8-p})$ is the volume form of the transverse $S^{8-p}$ and
 \eqn\solua{\eqalign{
 &h^2=1+\left({\rho_p\over\rho}\right)^{7-p}\ , \qquad
 f\equiv 1+g=1-\left({\rho_H\over\rho}\right)^{7-p} \ ,\cr
  e^{2 \Phi_{\infty}}Q_p^2=&(7-p)^2\rho_p^{7-p}\left(\rho_p^{7-p}+\rho_H^{7-p}\right) \ .
 }}
The non-extremality parameter is $\rho_H$, and by taking it to zero we obtain the extremal solution.
The symmetry group of the metric is $ISO(p)\times SO(9-p)$ ($p\neq 3$), and is enhanced to
 $ISO(1,p)\times SO(9-p)$ in the extremal case.
 $ISO(1,p)$ ($ISO(p)$) is the Poincare symmetry group of the D-brane worldvolume, and $SO(9-p)$ is the rotational symmetry in
 the $(9-p)$ transverse directions.
 In the following we will use these solutions as starting points in order to generate the non-relativistic Dp-brane backgrounds.

\subsec{Non-relativistic Dp-branes}
\subseclab\procedures

The non-relativistic Dp-brane backgrounds can be obtained from the relativistic Dp-brane solutions
via a TsT transformation as in \MaldacenaWH,
or by using the Null Melvin Twist procedure as in \HerzogWG\AdamsWT, following \AlishahihaRU\GimonXK.
In the case of the type IIB Dp-branes it works as in the original example of \HerzogWG\AdamsWT\MaldacenaWH.
We will verify that both methods generate the same non-relativistic Dp-brane backgrounds.

Starting with the black brane solutions \solu, the procedure of \MaldacenaWH\ consists
of the following steps:
\item{(1)} T-dualize along an isometry direction $\chi$
in the compact directions. In the type IIB case, the odd-dimensional spheres can be
written as $U(1)$ fibrations over a complex projective space as
$U(1)\hookrightarrow S^{8-p}\hookrightarrow \PP^{7-p\over2} $ and we T-dualize
along the fiber direction $d\chi$ (details of the relevant fibrations are outlined
in the Appendix).
\item{(2)} Shift a light cone direction in the world-volume of the brane according
to $\xi={1\over\sqrt{2}}(y-\tau)\to\xi+\delta\chi$.

\item{(3)} Repeat (1).

The Null Melvin Twist generates new supergravity solutions in eight steps, and consists of a boost, a TsT transformation and an inverse boost plus a scaling limit:

\item{1)} Start with the black brane solutions in \solu.

\item{2)} Boost in the translationally invariant direction $y$ by an amount $\gamma$.

\item{3)} T-dualize along the boosted direction $dy$.

\item{4)} Twist a one form in the transverse compact direction $d\chi\to d\chi+\alpha dy$.

\item{5)} T-dualize back along $dy$.

\item{6)} Boost back by $-\gamma$ along $y$.

\item{7)} Take the scaling limit $\alpha\to0$, $\gamma\to\infty$ keeping $\beta=\half\alpha e^\gamma$ fixed.

\item{8)} Change to lightcone coordinates $t=(y+\tau)/\sqrt{2}$, $\xi=(y-\tau)/\sqrt{2}$.

\noindent By applying this procedure, we obtain the non-relativistic black Dp-branes (for odd $p$)
 \eqn\lightsol{\eqalign{
 ds^2=&{1\over hK}\left[-\left({g\over2}+{2\beta^2\rho^2f}\right)dt^2-{g\over2}d\xi^2+(1+f)dtd\xi+K(dx^i)^2\right] \cr
 &+h\left[f^{-1}d\rho^2+\rho^2\left({1\over K}(d\chi+{\cal A})^2+ds^2_{\PP}\right)\right] \ ,\cr
 B=&{\beta\rho^2\over\sqrt{2}K}(d\chi+{\cal A})\wedge\left((1+f)dt+(1-f)d\xi\right)\ ,\qquad e^\Phi=\sqrt{h^{3-p}\over K}\ ,\cr
 }}
where $K=1-\beta^2\rho^2 g(\rho)$ and $g(\rho)$ is defined in \solua.\foot{In the procedure of \MaldacenaWH, the shift $\delta^2 = 2 \beta^2$.}
We introduced
$d{\cal A}=J$, where $J$ is the Kahler form on the complex projective space $\PP^{7-p\over2}$.
$ds^2_\PP$ is the metric on $\PP^{7-p\over2}$
and
$$
{1\over K}(d\chi+{\cal A})^2+ds^2_{\PP} \ ,
$$
is the metric on the squashed $(8-p)$-sphere, with squashing parameter $K$
(the details are in the Appendix). Note that the RR flux is not affected by the Null Melvin Twist. When $p=3$ the solution reduces to the known one \AdamsWT.

The Poincare symmetry group of the relativistic Dp-brane metric is replaced by
a non-relativistic symmetry. Asymptotically, at large $\rho$ it includes space and time translations, space rotations and Galilean transformations.
The extremal non-relativistic Dp-brane metric exhibits asympotically also the $SO(9-p)$ rotational symmetry group.
This rotational symmetry is broken in the non-extremal case to $SU({9-p\over 2}) \times U(1)$, since the odd-dimensional spheres get squashed and the remaining isometries are the ones of the complex projective space $\PP^{7-p\over2}$ times the fiber direction.

Next, we take the near--horizon limit by replacing the harmonic function $h\to\left({\rho_p\over\rho}\right)^{7-p\over2}$ and changing the radial coordinate $\rho/\rho_p=\rho_p/r$. The near--horizon geometry of the non-relativistic non-extremal black branes reads
 \eqn\lightsol{\eqalign{
 ds^2=&\left({\rho_p\over r}\right)^{7-p\over2}{1\over K}\left[-\left({g\over2}+{2f\Delta^2\over r^2}\right)dt^2-{g\over2}d\xi^2+(1+f)dtd\xi+K(dx^i)^2\right] \cr
 &+\left({\rho_p\over r}\right)^{p+1\over2}\left[f^{-1}dr^2+r^2\left({1\over K}(d\chi+{\cal A})^2+ds^2_\PP\right)\right] \ ,\cr
 B=&{\Delta\rho_p^2\over \sqrt{2}r^2K}(d\chi+{\cal A})\wedge\left((1+f)dt+(1-f)d\xi\right)\ ,\cr
 e^\Phi=&{1\over\sqrt{K}}\left({\rho_p\over r}\right)^{{(p-3)(7-p)\over4}} \ , }}
where the RR charge \solua\ is replaced by $Q_p\to(7-p)\rho_p^{7-p}$, we introduced the notations $\Delta=\beta\rho_p^2$, $r_H=\rho_p^2/\rho_H$, and
 \eqn\wheree{
 f\equiv1+g=1-\left({r\over r_H}\right)^{7-p}\ ,\qquad
 K=1+{\Delta^2 r^{5-p}\over r_H^{7-p}}\ .
 }

The Hawking temperature of the non-relativistic black branes can be computed
in the usual way by requiring the absence of a conical singularity of the
euclideanized metric near the horizon at $r=r_H$. One finds
\eqn\HT{
T_H={\sqrt{2}(7-p)\over 4\pi r_H}\left({\rho_A\over r_H}\right)^{{3-p\over 2}}\,.}
The factor of $\sqrt{2}$ is due to the relation between the time coordinate
which is used in the near horizon analysis, and the time coordinate of the
non-relativistic field theory \AdamsWT.

Finally, we can take the extremal limit $r_H\to\infty$, where $f\to1$, $g\to0$, $K\to1$ and find the zero-temperature solution with non-relativistic symmetries
 \eqn\solschre{\eqalign{
 ds^2=&\left({\rho_p\over r}\right)^{7-p\over2}\left[-{2\Delta^2\over r^2}dt^2+2dtd\xi+(dx^i)^2\right] +\left({\rho_p\over r}\right)^{p+1\over2}\left[dr^2+r^2\left((d\chi+{\cal A})^2+ds^2_{\PP}\right)\right] \ ,\cr
 B=&{\sqrt{2}\Delta\rho_p^2\over r^2}(d\chi+{\cal A})\wedge dt\ ,\qquad e^\Phi=\left({\rho_p\over r}\right)^{{(p-3)(7-p)\over4}}\ .
 }}
 The parameter $\Delta$ can be eliminated by a redefinition of the coordinates $t$ and $\xi$.
 Note, however, that this cannot be done for the non-extremal Dp-brane solutions and $\Delta$
 should be a physical parameter of the dual non-relativistic field theory related to a chemical potential \AdamsWT.
 Also,  the $SO(9-p)$ rotational symmetry group
which was broken in the non-extremal case is restored in the extremal limit.

\subsec{Non-relativistic NS-branes}

Consider the near--horizon geometry \lightsol\ for the black Dp-branes. Under an S-duality transformation we have
 \eqn\sdual{
 \eqalign{\Phi'=-\Phi\ , \qquad& ds'^2=e^{-\Phi}ds^2 \ ,\cr
 B_2'=C_2\ , \qquad& C_2'=-B_2 \ ,
 }}
where $C_2$ is the RR two-form potential. We can obtain the NS fivebrane and the fundamental string solutions for $p=5$ and $p=1$.

{\it Non-relativistic NS5-branes}

S-duality transformation of the non-relativistic D5-branes yields the
type IIB non-relativistic NS5-brane  background
\eqn\nsfivetemp{
\eqalign{
ds_{NS5}^2=&{1\over\sqrt{K}}\left[-\left({g\over2}+{2f\Delta^2\over r^2}\right)dt^2-{g\over2}d\xi^2+(1+f)dtd\xi+K(dx^i)^2\right] \cr &+\sqrt{K}\left({\rho_5\over r}\right)^2\left[f^{-1}dr^2+r^2\left({1\over K}(d\chi+{\cal A})^2+ds^2_{\PP\,{}^1}\right)\right] \ ,\cr
H_3=&Q_5{\rm Vol}(S^3) ,\qquad e^{\Phi}={r\over\rho_5}\sqrt{K} \ ,\cr
 C_2=&-{\Delta\rho_5^2\over \sqrt{2}r^2K}(d\chi+{\cal A})\wedge\left((1+f)dt+(1-f)d\xi\right)\ .\cr
 }}
The metric exhibits asymptotically the same symmetries as the non-relativistic D5-brane metric. In the limit $r_H\to\infty$ we find the zero-temperature non-relativistic geometry
\eqn\nsfive{
\eqalign{
ds_{NS5}^2=&-{2\Delta^2\over r^2}dt^2+2dtd\xi+(dx^i)^2+\left({\rho_5\over r}\right)^2\left(dr^2+r^2[(d\chi+{\cal A})^2+ds^2_{\PP\,{}^1}]\right)
\ ,\cr
C_2=&-{\sqrt{2}\Delta\rho_5^2\over r^2}(d\chi+{\cal A})\wedge dt \ ,\qquad e^{\Phi}={r\over\rho_5} \ ,
}}
and the NSNS flux remains unchanged. This is a background with an NS5 and a D1 string stretching along the $U(1)$ fiber direction $d\chi+{\cal A}$. Note that this is different from the solution one would get by starting with the non-extremal NS5 brane and applying the Null Melvin Twist.

{\it Non-relativistic fundamental string}

S-duality transformation of the non-relativistic D1-branes yields the
non-relativistic type IIB fundamental string background
\eqn\fonetemp{
\eqalign{
ds_{F1}^2=&{1\over\sqrt{K}}\left({\rho_1\over r}\right)^6\left[-\left({g\over2}+{2f\Delta^2\over r^2}\right)dt^2-{g\over2}d\xi^2+(1+f)dtd\xi\right] \cr
&+\sqrt{K}\left({\rho_1\over r}\right)^4\left[f^{-1}dr^2+r^2\left({1\over K}(d\chi+{\cal A})^2+ds^2_{\PP\,{}^3}\right)\right] \ ,\cr
\star H_3=&{\rm Vol}(S^7)Q_{1} \ ,\qquad e^{\Phi}=\sqrt{K}\left({\rho_1\over r}\right)^3 \ ,\cr
C_2=&-{\sqrt{2}\Delta\rho_1^2\over r^2}(d\chi+{\cal A})\wedge dt \ .
}}
In the limit $r_H\to\infty$ we find
\eqn\fonebis{
\eqalign{
ds_{F1}^2=&\left({\rho_1\over r}\right)^6\left[-{2\Delta^2\over r^2}dt^2+2dtd\xi\right]+\left({\rho_1\over r}\right)^4\left(dr^2+r^2[d(\chi+{\cal A})^2+ds^2_{{\PP\,{}^3}}]\right) \ ,\cr
C_2=&-{\sqrt{2}\Delta\rho_1^2\over r^2}(d\chi+{\cal A})\wedge dt \  ,\qquad e^{\Phi}=\left({\rho_1\over r}\right)^3 \ ,
}}
while the NSNS flux is unchanged. This is a background with a fundamental string and a D1 string stretching along the $U(1)$ fiber direction $d\chi+{\cal A}$. Again, this solution is different from the one obtained by starting with the non-extremal F1 and applying the Null Melvin Twist procedure.

{\it Non-relativistic D3-branes and S-duality}

Unlike the relativistic case, the non-relativistic D3-brane of type IIB is not invariant under S-duality, since the NSNS potential is mapped
to a RR 2-form potential $C_2$. This implies that unlike ${\cal N}=4$ Super Yang-Mills, here the dual non-relativistic conformal field theory does not possess
an S-duality symmetry and moreover it is dual to a purely RR background. For large values of the dilaton, the correct description of the non-relativistic D3-brane is in terms of the dual background
\eqn\sdualthree{
\eqalign{
ds_{D3'}=&\left({\rho_3\over r}\right)^2\sqrt{K}\left[-\left({g\over2}+{2f\Delta^2\over r^2}\right)dt^2-{g\over2}d\xi^2+(1+f)dtd\xi+K(dx^i)^2\right] \cr
 &+\left({\rho_3\over r}\right)^{2}\sqrt{K}\left[f^{-1}dr^2+r^2\left({1\over K}(d\chi+{\cal A})^2+ds^2_{\PP\,{}^2}\right)\right] \ ,\cr
 C_2=&-{\Delta\rho_3^2\over \sqrt{2}r^2K}(d\chi+{\cal A})\wedge\left((1+f)dt+(1-f)d\xi\right)\ ,\cr
 F_5=&Q_3(1+\star){\rm Vol}(S^5)\ ,\qquad e^\Phi=\sqrt{K}\ .}}

\newsec{Non-relativistic Type IIA Dp-branes}

In this section, we will first perform a Null Melvin Twist of the type IIA black branes and obtain a warped non-relativistic metric whose noncompact part depends explicitly on the coordinates of the sphere.
We will then modify the solution and get a solution analogous to the type IIB case, where the noncompact part of the metric is independent of the compact directions.

Consider the black brane solution \solu. In the case of even $p$, the compact directions transverse to the brane consist of even-dimensional spheres, that cannot be written as $U(1)$ fibrations. Let us consider the extremal case for simplicity and perform a Null Melvin Twist, where the shift in the TsT is along one of the isometries of the compact sphere. If we take the standard metric on the sphere
 \eqn\metricsphere{
 d\Omega_{8-p}=d\theta_1^2+\sin^2\theta_1[d\theta_2^2+\sin^2\theta_2(\ldots +\sin^2\theta_{8-p-1}d\theta_{8-p})] \ ,
 }
and choose  $d\theta_{8-p}$ as the isometry direction, we get the near--horizon geometry
 \eqn\schroangle{\eqalign{
  ds^2=&\left({\rho_p\over r}\right)^{7-p\over2}{1\over K}\left[-{2f\Delta^2e(\theta)\over r^2}dt^2+2dtd\xi+(dx^i)^2\right] +\left({\rho_p\over r}\right)^{p+1\over2}\left[dr^2+r^2d\tilde\Omega^2_{8-p}\right] \ ,\cr
 B=&{\sqrt{2}\Delta\rho_p^2e(\theta)\over r^2}d\theta_{8-p}\wedge dt\ ,\qquad e^\Phi=\left({r\over \rho_p}\right)^{(3-p)(7-p)\over4}\ ,  \cr
  \star F_{p+2}=&{\rm Vol}(S^{8-p})Q_p \ .
}}
The components of the metric parallel to the brane explicitly depend on the angles of the sphere through the function
$ e(\theta)=\sin^2\theta_1\ldots\sin^2\theta_{8-p-1}$.
However, we can drop the angle dependence from $g_{tt}$ and find a modified $B$ field whose stress tensor satisfies the new equations of motion. It is straightforward to check that the following solution has the desired properties of a non-relativistic extremal geometry
 \eqn\schroiia{\eqalign{
 ds^2=&\left({\rho_p\over r}\right)^{7-p\over2}\left[-{2\Delta^2\over r^2}dt^2+2dtd\xi+(dx^i)^2\right] +\left({\rho_p\over r}\right)^{p+1\over2}\left[dr^2+r^2d\Omega^2_{8-p}\right] \ ,\cr
 B=&{\sqrt{9-p}\Delta\rho_p^2e(\theta)\over r^2}d\theta_{8-p}\wedge dt\ ,
 }}
with the dilaton and RR flux as in \schroangle. The metric in the directions parallel to the brane does not depend on the compact directions, and has the same
structure as that of the odd $p$ non-relativistic Dp-branes.
We leave it as open problem to construct the finite temperature version of
\schroiia.

\newsec{Non-relativistic M-theory branes}

In this section, we will obtain the extremal solutions for the non-relativistic M theory geometries of the near-horizon M2 and M5 branes and the KK monopole. The M2 and M5 solutions exhibit the full Schr\"{o}dinger symmetry with dynamical exponents $\nu=1/2$ and $\nu=2$ respectively, as opposed to the D3 brane case \HerzogWG\AdamsWT\MaldacenaWH\ that has $\nu=1$. By reducing to ten dimensions we will then obtain the type IIA NS5-brane and F1 solutions.

The uplift to M theory of the IIA solutions \schroiia\ is given by
\eqn\lift{\eqalign{
ds^2_{11}=&e^{-{2\over3}\Phi}ds_{IIA}^2+e^{{4\over3}\Phi}\left(dx_{10}^2+dx^\mu C_\mu\right)^2 \ ,\cr
A_3=&{1\over3!}dx^\mu\wedge dx^\nu\wedge dx^\rho C_{\mu\nu\rho}+\half dx^\mu\wedge dx^\nu\wedge dx_{10}B_{\mu\nu} \ ,
}}
where $x_{10}$ denotes the eleventh coordinate, $C$'s are the RR potentials and $B$ is the NSNS potential.

\subsec{Non-relativistic M2-branes}

The uplift of the D2 solution \schroiia\ is most transparent in cartesian coordinates. The non-relativistic D2-brane reads
 \eqn\cartedtwo{
 \eqalign{
 ds^2_{D2}=&\left({\rho\over \rho_2}\right)^{5\over2}\left(-2\beta^2\rho^2dt^2+2dtd\xi+dx^2\right)+
\left({\rho_2\over \rho}\right)^{5\over2}\left(dx_3^2+\ldots+dx_9^2\right) \ ,\cr
B=&\sqrt{7}\beta\left(x_8dx_9-x_9dx_8\right)\wedge dt \ ,\qquad e^\Phi=\left(\rho_2/\rho\right)^{5\over4} \ ,\cr
F_4=&{1\over H_2^2}{Q_2\over\rho^7}dt\wedge d\xi\wedge dx\wedge\left(x_3dx_3+\ldots+x_9dx_9\right) \ ,
}}
where $\rho^2=x_3^2+\ldots+x_9^2$. Its lift to eleven dimensions is
 \eqn\uplifttwo{
 \eqalign{
 ds^2=&H^{-{2\over3}}\left(-2\beta^2\rho^2dt^2+2dtd\xi+dx^2\right)+
H^{1\over3}\left(dx_3^2+\ldots+dx_9^2+dx_{10}^2\right) \ ,\cr
F_4=&{1\over H_2^2}{Q_2\over\rho^7}dt\wedge d\xi\wedge dx\wedge\left(x_3dx_3+\ldots+x_9dx_9\right) +2\sqrt{7}\beta dt\wedge dx_8\wedge dx_9\wedge dx_{10} \ ,
}}
where $x_{10}$ is the M theory circle and $H=\left({\rho_2\over \rho}\right)^5$ is the harmonic function corresponding to a smeared M2-brane and $F_4$ is the eleven dimensional four-form flux. To obtain the localized M2-brane \ItzhakiDD\ one needs to replace the harmonic function $H$ with
 \eqn\newharm{
 H\to\tilde H=\sum_{n=-\infty}^\infty {\tilde \rho_2^6\over [\rho^2+(x_{10}-\bar x_{10}+2\pi n R)^2]^3} \ ,
 }
where $\bar x_{10}$ is the position of the M2-brane on the eleventh dimensional circle of radius $R$. In the limit where we are far away from the M2-brane, one can Poisson resum \newharm\ and get back to the smeared solution, since
$$\eqalign{
\tilde H={\rho_2^5\over\rho^5}+\sum_{m=1}^\infty e^{-m\rho/R}{\cal O}\left({\rho_2^5\over \rho^5}\right)\ ,\cr
}$$
where we matched
$\tilde \rho_2^6\propto R\rho_2^5$.
In the deep IR we are very close to the M2-brane and we can neglect all the images, namely take $\tilde H=(\tilde \rho_2/\tilde\rho)^6$,  with $\tilde\rho^2=\rho^2+x_{10}^2$, obtaining the non-compact eleven dimensional solution
\eqn\mtwobis{
\eqalign{
 ds^2_{M2}=&\tilde H^{-{2\over3}}\left(-2\beta^2\tilde \rho^2dt^2+2dtd\xi+dx^2\right)+
\tilde H^{1\over3}\left(dx_3^2+\ldots+dx_9^2+dx_{10}^2\right) \ ,\cr
F_4=&{1\over\tilde H^2}{\tilde Q_2\over \tilde \rho^8}dt\wedge d\xi\wedge dx\wedge\left(x_3dx_3+\ldots+x_9dx_9\right) +2\sqrt{8}\beta dt\wedge dx_8\wedge dx_9\wedge dx_{10} \ ,
}}
where we replaced $\sqrt{9-p}\to\sqrt{10-p}$ as coefficient of the last term in the four-form flux. Finally we change variables to $\tilde\rho^2/\tilde \rho_2^2=\tilde\rho_2/r$ and rescale $r\to r/2$, $\tilde \rho_2\to \tilde \rho_2/2$, we obtain the non-relativistic geometry
\eqn\mtwo{
\eqalign{
 ds^2_{M2}=&\left({\tilde\rho_2\over r}\right)^2\left(-2{\tilde\Delta_2\over r}dt^2+2dtd\xi+dx^2+dr^2\right)+
4\tilde \rho_2^2d\Omega_7^2\ ,
}}
where $\tilde\Delta_2=4\beta^2 \tilde\rho_2^3$.
This solution exhibits the full Schr\"{o}dinger symmetry with dynamical exponent
 $$
 \nu=\half \ .
 $$
Note that unlike the non-relativistic D3-brane case, special conformal transformation is not a symmetry of the non-relativistic M2-brane metric.

\subsec{Non-relativistic M5-branes}

Consider the solution for the type IIA D4-brane \schroiia. Its lift to eleven dimensions, after changing the radial coordinate to $\rho/\rho_4=\rho_4/r$ and rescaling $r\to 2r$, $\rho_4\to 2\rho_4$, reads
\eqn\liftdfour{\eqalign{
 ds^2_{M5}=&\left({\rho_4\over r}\right)^{2}\left[-{2\tilde\Delta_5^4\over r^4}dt^2+2dtd\xi+(dx^i)^2+dr^2\right] +{1\over4}\rho_4^2d\Omega_4^2 \ ,\cr
 F_4=&2\sqrt{5}\beta dt\wedge dx_8\wedge dx_9\wedge dx_{10}+
 Q_4 {\rm Vol}(S^4) \ ,
 }}
where $i=1,\ldots,4$ and the last coordinate is the eleventh dimensional one and $\tilde \Delta_5^2=\beta\rho_4^3/2$. This solution exhibits the full Schr\"{o}dinger symmetry with dynamical exponent
 $$
 \nu=2 \ .
 $$
As for the non-relativistic M2-branes, also the M5-brane metric does not possess the special conformal symmetry.

\subsec{Non-relativistic KK monopole}

The D6-brane in \schroiia\ is uplifted to an eleventh dimensional solution
\eqn\dsix{
\eqalign{
ds^2_{KK}=&-{2\Delta^2\over r^2}dt^2+2dtd\xi+(dx^i)^2\cr
&+\left({\rho_6\over r}\right)^3\left[dr^2+r^2[d\theta^2+\sin^2\theta d\varphi^2+{1\over \rho_6^2}\left(dx_{10}^2-Q(\cos\theta-1)d\varphi\right)^2]\right] \ ,\cr
A_3=&{\sqrt{3}\beta\rho_6^4\over r^2}\sin^2\theta_1\ d\varphi\wedge dt\wedge dx_{10} \ ,
 }}
We can identify a non-relativistic KK monopole, whose compact part is an $ALE_{N-1}$ space, by changing coordinates to
$$\eqalign{
\tilde\theta=\theta/2\ ,\qquad&\tilde \psi=x_{10}/2\rho_6 \ ,\cr
\tilde \varphi=\varphi+\tilde \phi\ , \qquad&r=4\tilde r^2/\rho_6 \ ,
\cr
}$$
and obtaining
 \eqn\KKnonrel{
 \eqalign{
ds^2_{KK}=&-{2\tilde\Delta_7^4\over \tilde r^4}dt^2+2dtd\xi+(dx^i)^2+\left({\rho_6\over \tilde r}\right)^4\left[d\tilde r^2+\tilde r^2[d\tilde \theta^2+\sin^2\tilde\theta d\tilde\varphi^2+\cos^2\tilde\theta d\tilde \psi^2]\right] \ ,\cr
F_4=&{\sqrt{3}\tilde\Delta_7^2\rho_6^4\over \tilde r^4}dt\wedge d\tilde\varphi\wedge d\tilde \psi\left(-{2d\tilde r\over \tilde r}\sin^2 2\tilde\theta+{\sin4\tilde\theta}d\tilde \theta\right) \ ,
}}
where we redefined $\tilde \Delta_7^2=\beta\rho_6^3/4$. The compact part is an $ALE_{N-1}$ space ($N=Q_6$), where we identified $(\tilde \varphi,\tilde \phi)\sim (\tilde \varphi,\tilde \phi)+(2\pi/N,2\pi/N)$. It is easy to see this by changing coordinates to $y={\rho_6^2\over\tilde r}$.

\subsec{Type IIA NS-branes}

{\it Non-relativistic NS5-branes}

Consider the M5-brane \liftdfour. By reducing it to ten dimensions along a transverse direction we obtain the solution for the IIA NS5-brane
 \eqn\iiafive{
 \eqalign{
 ds^2_{NS5}=& -2\beta^2\rho^2 dt^2+2dtd\xi+ (dx_i)^2+(\rho_4^3/\rho^3)\left(dx_6^2+\ldots+dx_9^2\right)\ ,\cr
 H_3=&Q_4{\rm Vol}(S^3) \ ,\qquad e^\Phi=\left(\rho_4/\rho\right)^{3\over2}\ , \cr
 F_4=&2\sqrt{5}\beta dt\wedge dx_4\wedge dx_8\wedge dx_9 \ .
 }}
By a change of variables we can recast this metric into the more familiar form
 \eqn\nsfiveiia{
 ds_{NS5}^2= -{2\Delta^2\over r^2 } dt^2+2dtd\xi+(dx_i)^2+{\rho_4\over r}\left(dr^2+r^2d\Omega_3^2\right)\ ,
 }
where $\Delta=\beta\rho_4$. While the type IIB fivebrane solution \nsfive\ contains D1-branes stretched along the $d\chi$ direction, the type IIA fivebrane solution contains D2-branes stretched along a worldvolume direction and a transverse direction.

{\it Non-relativistic fundamental string}

Reducing the M2-brane solution to ten dimensions we can obtain the IIA fundamental string solution with non-relativistic symmetries
 \eqn\iiafund{
 \eqalign{
    ds^2_{F1}=&{\rho^6\over \tilde \rho_2^6}\left(-2\beta\rho^2dt^2+2dtd\xi\right)+dx_2^2+\ldots+dx_{9}^2 \ ,\cr
    F_4=&2\sqrt{7}\beta dt\wedge dx_7\wedge dx_8\wedge dx_9 \ ,\cr
    H_3=&-{\rho^4\over\tilde\rho_2^{12}} dt\wedge d\xi\wedge \left(x_2dx_2+\ldots+x_9dx_9\right) \ ,\qquad e^\Phi=(\rho/ \tilde\rho_2)^3 \ ,
    }}
By a change of coordinates we can recast the metric in the following form
 \eqn\iiafundbis{
 ds_{F1}^2=\left({\tilde \rho_2\over r}\right)^6\left(-{2\Delta^2\over r^2}dt^2+2dtd\xi\right)+\left({\tilde\rho_2\over r}\right)^4\left(dr^2+r^2d\Omega_7^2\right) \ ,
 }
where $\Delta=\beta\tilde\rho_2$.

\newsec{Non-relativistic intersecting branes}

\subsec{Two D1-D5 non-relativistic systems}

We can apply a TsT transformation to the black D1-D5 system in \HorowitzAY. The four compact directions parameterize a four manifold $M_4$ that can be either a K3, that has no isometries, or a torus $T^4$. In both cases we can perform a T-duality  along the three-sphere transverse to the D1, by considering the sphere as a Hopf fibration and proceeding as above. However, when the compact four-manifold is a torus, we can perform a TsT transformation with a T-duality along the torus direction. In this case, we obtain a new solution with an extra five-form flux turned on, whose metric is the same as the previous one
in the limit of zero temperature. Hence, in the case that the compact four-manifold is a torus, we have two different non-relativistic
black branes that have the same metric in the zero temperature limit In fact, there is a continuous family of non-relativistic backgrounds, all
having the same metric asymptotics. They are obtained
by a T-duality along a linear combination of a $T^4$ direction and the $U(1)$
fiber of $S^3$.
It would be interesting to understand this result in the dual non-relativistic quantum mechanics.

Starting with the Schwarzschild black brane geometry in \HorowitzAY\ (with the boost charge $\sigma$ set to zero)
 \eqn\maldahoro{
 \eqalign{
 ds^2_{D1D5}=&{1\over \sqrt{H_1H_5}}\left[ - fd\tau^2+dy^2\right]
+\sqrt{H_1H_5}\left[{d\rho^2\over f}+\rho^2\left({(d\chi+{\cal A})^2}+d\Omega_{\PP\,{}^1}\right)\right]+\sqrt{H_1\over H_5}ds^2_{M_4} \,\cr
F_3=&{Q_1\over H_1^2}{1\over \rho^3}d\tau\wedge dy\wedge d\rho+Q_5{\rm Vol}(S^3) \ ,  \qquad e^{\Phi}=\sqrt{H_1\over H_5 } \ ,
 }}
where the compact manifold is either a K3 or a torus, we can perform a TsT transformation with T-duality along the sphere. The non-relativistic D1-D5 solution is similar to the ones constructed above for the odd branes
\eqn\blackdonedfive{\eqalign{
ds^2_{D1D5}=&{1\over \sqrt{H_1H_5}K}\left[-\left({g\over2}+2\rho^2\beta^2 f\right)dt^2-{g\over2}d\xi^2+(1+f)dtd\xi\right]\cr
&+\sqrt{H_1H_5}\left[{d\rho^2\over f}+\rho^2\left({(d\chi+{\cal A})^2\over K}+d\Omega_{\PP\,{}^1}\right)\right]+\sqrt{H_1\over H_5}ds^2_{M_4} \ ,\cr
B=&{\beta\rho^2\over\sqrt{2}}{1\over K}(d\chi+{\cal A})\wedge\left((1+f)dt+(1-f)d\xi\right)\ ,\qquad e^{\Phi}=\sqrt{H_1\over H_5 K} \ ,\cr
F_3=&{Q_1\over H_1^2}{1\over \rho^3}dt\wedge d\xi\wedge d\rho+Q_5{\rm Vol}(S^3) \ ,
}}
where the harmonic functions are $H_1=1+\rho_1^2/\rho^2$, $H_5=1+\rho_5^2/ \rho^2$ and $K=1+\beta^2\rho_H^2$ is a constant, $f=1+g$ and $g=-\rho_H^2/\rho^2$. The black brane charges are $Q_1=2\rho_1\sqrt{\rho_1^2+\rho_H^2}$, $Q_5=2\rho_5\sqrt{\rho_5^2+\rho_H^2}$. The near horizon Schwarzschild geometry is
\eqn\blacknear{\eqalign{
ds^2_{D1D5}=&{\rho_1\rho_5\over r^2K}\left[-\left({g\over2}+{2\Delta^2 f\over r^2}\right)dt^2-{g\over2}d\xi^2+(1+f)dtd\xi+{K\over f}dr^2\right]\cr
&+\rho_1\rho_5\left({(d\chi+{\cal A})^2\over K}+d\Omega_{\PP\,{}^1}\right)+{\rho_1\over \rho_5}ds^2_{M_4} \ ,\cr
B=&{\Delta\rho_1\rho_5\over\sqrt{2}r^2}{1\over K}(d\chi+{\cal A})\wedge\left((1+f)dt+(1-f)d\xi\right)\ ,\qquad e^{\Phi}={\rho_1\over\rho_5\sqrt{K}}\ ,\cr
F_3=&2\rho_5^2\left(-{1\over r^3}dt\wedge d\xi\wedge d r+{\rm Vol}(S^3)\right) \ ,
}}
where $\Delta=\beta\rho_1\rho_5$. We changed variables to ${\rho^2\over \rho_1\rho_5}={\rho_1\rho_5\over r^2}$ and we replaced $Q_1\to2\rho_1^2$, $Q_5\to2\rho_5^2$ and the RR flux is the same as in \blackdonedfive, while $g=-r^2/r_H^2$ where $r_H=\rho_1\rho_5/\rho_H$. In the limit of zero temperature the solution becomes
\eqn\zerotemp{\eqalign{
ds^2_{D1D5}=&{\rho_1\rho_5\over r^2}\left[-\left(
{2\Delta^2 \over r^2}\right)dt^2+2dtd\xi+dr^2\right]+\rho_1\rho_5\left((d\chi+{\cal A})^2+d\Omega_{\PP\,{}^1}\right)+{\rho_1\over \rho_5}ds^2_{M_4} \ ,\cr
B=&{\sqrt{2}\Delta\rho_1\rho_5\over r^2}(d\chi+{\cal A})\wedge dt\ ,\qquad e^{\Phi}={\rho_1\over\rho_5}\ ,
}}
plus the usual $F_3$ flux. Its dual is a $(0+1)$--dimensional quantum mechanics with Schr\"{o}dinger symmetry and exponent $\nu=1$. Note that this solution has vanishing scalar curvature.
 For large values of the dilaton, we can go to the S-dual solution by applying \sdual.

The other possibility is to
perform the T-duality along a torus direction, instead of using the $U(1)$ isometry of the Hopf fibration. We denote by $\varphi_i, i=1,...,4$ the torus coordinates and we pick $d\varphi_4$ as the T-duality isometry direction. Following the TsT method of \MaldacenaWH\ we find the following near-horizon black hole solution\foot{The Null Melvin Twist analogous to the TsT just described would be to T-dualize the $dy$ direction, shift the torus isometry $d\varphi_4$ and then T-dualize back, as outlined in section \procedures. However, in performing the scaling limit some of the fluxes blow up, so in this case the Null Melvin Twist and the TsT transformation are not equivalent.}
 \eqn\newtorus{
 \eqalign{
 ds^2_{D1D5}=&{\rho_1\rho_5\over r^2}{1\over K}\left[-\left({g\over2}+{2\Delta f\over r^2}\right)dt^2-{g\over2}d\xi^2+(1+f)dtd\xi+{K\over f}dr^2\right]\cr
 &+\rho_1\rho_5d\Omega_3^2+{\rho_1\over \rho_5}\left(\sum_{i=1}^3d\varphi_i^2+d\varphi_4^2/K\right) \ ,\cr
 B=&-{\Delta\rho_1\over \sqrt{2} Kr^2}d\varphi_4\wedge\left[(1+f)dt+(1-f)d\xi\right] \ ,\qquad e^\Phi={\rho_1\over \rho_5 \sqrt{K}} \ ,\cr
 F_5=&{\Delta\rho_1\rho_5^2\over\sqrt{2} Kr^2} \Bigl[-{\cos\theta-1\over r}dt\wedge dr\wedge d\chi\wedge d\phi\cr
 &+(1+2\,\star_{10})\left[(1+f)dt+(1-f)d\xi\right]\wedge{\rm Vol}(S^3)\Bigl]\wedge d\varphi_4 \ ,
 }}
where $\Delta=\alpha\rho_1/\sqrt{2}$ ($\alpha$ is the shift) and the three-form $F_3$ is the same as in \blacknear. The self-dual flux in the type IIB equations of motion is the combination $\tilde F_5=F_5-\half (B\wedge F_3-C_2\wedge H)$. The fluxes in \newtorus\ conspire to make a self dual $\tilde F_5$, which is the reason why we have a factor $(1+2\,\star)$ instead of the usual $(1+\star)$.
In the extremal limit, this second solution has the same metric, dilaton and three-form  as in \zerotemp, but a different B-field, plus a five-form flux
\eqn\differen{\eqalign{
B=&-{\sqrt{2}\Delta\rho_1\over r^2}d\varphi_4\wedge dt \ ,\cr
F_5=&{\Delta\rho_1\rho_5^2\over\sqrt{2} r^2} \Bigl[-{\cos\theta-1\over r}dt\wedge dr\wedge d\chi\wedge d\phi+2(1+2\,\star_{10})\,dt\wedge{\rm Vol}(S^3)\Bigl]\wedge d\varphi_4 \ .
 }}

\subsec{Two F1-NS5 non-relativistic systems}

When the dilaton is large, a proper description of the non-relativistic D1-D5 system is obtained by applying the S-duality transformation \sdual\ that gives a non-relativistic F1-NS5 system with flux. In the case the T-duality in the TsT is along the three-sphere isometry, the near horizon geometry of such F1-NS5 solution at non-zero temperature is
 \eqn\fonensfive{\eqalign{
ds^2_{F1NS5}=&{\rho_5^2\over r^2\sqrt{K}}\left[-\left({g\over2}+{2\Delta^2 f\over r^2}\right)dt^2-{g\over2}d\xi^2+(1+f)dtd\xi+{K\over f}dr^2\right]\cr
&+\rho_5^2\sqrt{K}\left({(d\chi+{\cal A})^2\over K}+d\Omega_{\PP\,{}^1}\right)+{\sqrt{K}}ds^2_{M_4} \ ,\cr
C_2=&-{\Delta\rho_1\rho_5\over\sqrt{2}r^2}{1\over K}(d\chi+{\cal A})\wedge\left((1+f)dt+(1-f)d\xi\right)\ ,\cr
B=&{\rho_5^2\over r^2}dt\wedge d\xi+{\rho_5^2\over 2}{(\cos\theta -1)}d\chi\wedge d\phi \ ,\qquad
e^{\Phi}={\rho_5\sqrt{K}\over\rho_1}\ .
}}
In the extremal limit, we simply have $K,f\to1$ and $g\to0$.

In the second case \newtorus, in which the T-duality in the TsT is performed along the torus direction, when the dilaton becomes large we can pass to the S-dual description in terms of the extremal F1-NS5 near-horizon geometry
 \eqn\FNStorus{
 \eqalign{
 ds^2_{F1NS5}=&{\rho_5^2\over r^2}{1\over K}\left[-\left({g\over2}+{2\Delta f\over r^2}\right)dt^2-{g\over2}d\xi^2+(1+f)dtd\xi+{K\over f}dr^2\right]\cr
 &+\rho_5^2d\Omega_3^2+\sqrt{K}\left(\sum_{i=1}^3d\varphi_i^2+d\varphi_4^2/K\right) \ ,\cr
 C_2=&{\Delta\rho_1\over \sqrt{2} Kr^2}d\varphi_4\wedge\left[(1+f)dt+(1-f)d\xi\right] \ ,
 }}
and same RR five-form flux as in \newtorus, while the dilaton and B field are the same as in \fonensfive.
Just as for the D1-D5 system, the zero temperature limit of the two F1-NS5 systems have the same metric isometries.

\newsec{Phase structure}

When trying to relate the gravity description to a dual non-relativistic field theory, we
need to compactify the null coordinate $\xi$.
The reason is, as noted before, that the momentum in the $\xi$ direction is interpreted as the
particle number of the non-relativistic
field theory, which is discrete.
However, a compact null-direction, which can be viewed as the limit of a small compact space-like circle,
takes us out of the validity of the supergravity approximation.
To properly analyze the phase structure of the non-relativistic Dp-branes we should consider the non-extremal
solutions because in the finite temperature case the null circle receives a finite
size with a conical singularity at the origin of the
radial direction. We can then trust the supergravity approximation away from the boundary.
In the following we will perform a simpler analysis and consider the extremal case, which is valid when the
coordinate $\xi$ is non-compact.
The phase structure that we find is very similar to that of the relativistic Dp-branes.
This, together with the fact that at finite temperature the horizon is not affected by a TsT transformation \HorowitzWT,
suggest that the phase structure of the non-extremal non-relativistic Dp-branes is similarly equivalent to that
of the non-extremal relativistic Dp-branes.

We define a dimensionless expansion parameter
\eqn\ex{
g_{eff}^2 = l_s^{-4} \rho_p^{p+1}r^{3-p} \ .
}
The curvature ${\cal R}$ associated with the non-relativistic Dp-brane metric takes the form
$$
l_s^2 {\cal R} \sim {1 / g_{eff}} \ ,
$$
which explicitly reads
\eqn\rsce{
{\cal R} \sim  \rho_p^{-{(p+1)/ 2}}r^{{(p-3) /2}} \ .
}
The effective string coupling takes the form
$$
 e^\Phi  \sim  {g_{eff}^{(7-p)/2} \over Q_p} \sim \left({\rho_p \over r}\right)^{{(7-p)(p-3) / 4}} \ ,
$$
where we used $Q_p \sim e^{-\Phi_{\infty}} ( \rho_p/l_s)^{7-p}$.

The supergravity action for the non-relativistic Dp-brane backgrounds scales as
$$
l_s^{-8} \int \sqrt{g} e^{-2 \Phi} {\cal R} d^{10}x \sim Q_p^2 \ .
$$
This suggests that the number of degrees of freedom of the corresponding non-relativistic field theory also scales as
$Q_p^2$, which is
the same as in the relativistic case.
Again, to properly perform the calculation we need to consider the finite temperature background.
However we expect, barring a possibility of a phase transition,  that the thermal non-relativistic field theory exhibits the same number of degrees
of freedom scaling.

The supergravity action for the non-relativistic M-brane backgrounds scales as
$$
l_p^{-9} \int \sqrt{g} {\cal R} d^{11}x \sim Q_p^n \ ,
$$
where $n={3\over 2}$ for the M2-branes and $n=3$ for the M5-branes.
Again, this suggests that that the number of degrees of freedom of the corresponding non-relativistic conformal field theory
scales as in the relativistic case.

The phase diagram of the non-relativistic branes in various dimensions can be analyzed as in the  relativistic brane
case \ItzhakiDD, and in the presence of a NSNS B-field \AlishahihaCI. It has a similar structure.
We will perform the analysis in the following and we will set the string scale to one, except when explicitly written.

\subsec{Non-relativistic D1-branes}

The dimensionless effective coupling  \ex\
 reads
$$
g_{eff}= l_s^{-2}\rho_1 r \ .
$$
We expect a perturbative non-relativistic quantum mechanics description when the effective coupling is small $g_{eff}\ll 1$.
This is the UV regime of length scales $r  \ll {1 / \rho_1}$. The description breaks down at
$g_{eff} \sim 1$, i.e. at a length scale $r  \sim {1 / \rho_1}$.
The curvature of the non-relativistic D1-brane metric \rsce\ is small at length scales  $r  \gg {1 / \rho_1}$.
The effective string coupling
$$
e^\Phi  \sim \left({r / \rho_1}\right)^3 \ ,
$$
is small at length scales $r  \ll \rho_1$. Thus, the type IIB supergravity description
\solschre\ is valid in the regime $ {1 / \rho_1} \ll r  \ll \rho_1$. At length scales $r \gg \rho_1$ the effective string coupling is large and
we need to apply an S-duality transformation to the non-relativistic D1-brane background.
This gives the non-relativistic fundamental string \fonebis.
The effective string coupling now is  $e^\Phi  \sim \left({\rho_1 / r}\right)^3$ and it vanishes in the IR at large r.
The curvature in the new string units ${\tilde l}_s$
$$
{\cal R} \sim {r^2 / \rho_1^4} \ ,
$$
blows up in the IR.
The type IIB supergravity description \fonebis\ breaks down at length scales $r \sim \rho_1^2$.
In the IR regime $r \gg \rho_1^2$ we expect a free non-relativistic conformal quantum mechanics description.

\ifig\figdone{The phase diagram of the D1-brane theory as a function of the inverse energy scale $r$. It flows from a perturbative non-relativistic non-conformal quantum mechanics in the UV to a free quantum mechanics in the deep IR.}
{\epsfxsize5in\epsfbox{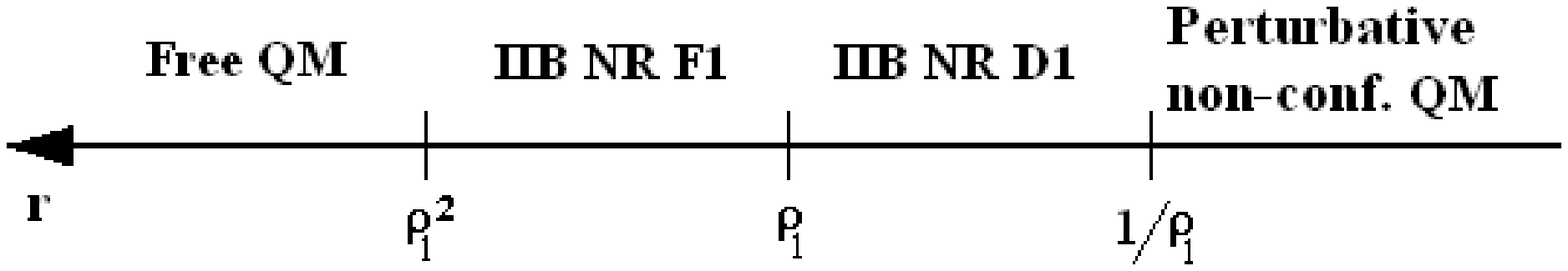}}

\subsec{Non-relativistic D2-branes}

The dimensionless effective coupling
$g_{eff}= l_s^{-2}\left(\rho_2^3 r\right)^{1 / 2}$ is small in the UV  at length scales $r  \ll {1 / \rho_2^3}$, and
we expect a perturbative  two-dimensional non-relativistic field theory description in this regime.
The description breaks down at a length scale $r  \sim {1 / \rho_2^3}$.
The curvature of the non-relativistic D2-brane metric \schroiia\ is small at length scales  $r  \gg {1 / \rho_2^3}$.
The effective string coupling
$e^\Phi  \sim \left({r / \rho_2}\right)^{{5/ 4}}$
is small at length scales $r  \ll \rho_2$. Thus, the type IIA supergravity description \schroiia\
is valid in the regime $ {1 / \rho_2^3} \ll  r  \ll \rho_2$. At length scales $r \gg \rho_2$ the effective string coupling is large and
we need to uplift the non-relativistic D2-brane ten-dimensional background to eleven dimensions.
The uplifted solution (see section 4.1) is valid as long as its curvature in eleven-dimensional Planck units
$ l_p^2 {\cal R} \sim e^{2 \Phi/3} l_s^2 {\cal R}$ is small. This gives $r \ll \rho_2^2$.
Thus the eleven-dimensional uplifted solution is valid in the regime  $\rho_2 \ll r \ll \rho_2^2$.
In the IR regime $r \gg \rho_2^2$ we expect a two-dimensional non-relativistic conformal field theory description
realized on the worldvolume of the
non-relativistic M2-branes.

\ifig\figone{The phase diagram of the D2-brane theory as a function of the inverse energy scale $r$. It flows from a perturbative non-relativistic non-conformal field theory in the UV to a non-relativistic CFT in the deep IR.}
{\epsfxsize5in\epsfbox{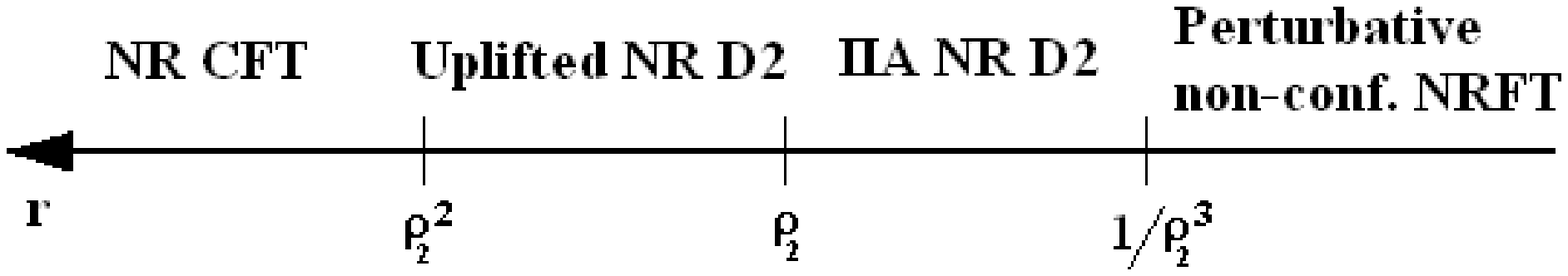}}

\subsec{Non-relativistic D4-branes}

The dimensionless effective coupling
$g_{eff}= l_s^{-2}\left({\rho_4^5 / r^2} \right)^{1 / 2}$ is small in the IR  at length scales $r  \gg \rho_4^{{5 / 2}}$, and
we expect a perturbative  four-dimensional non-relativistic field theory description in this regime.
The description breaks down at a length scale $r  \sim  \rho_4^{{5 / 2}}$.
The curvature of the non-relativistic D4-brane metric \schroiia\ is small at length scales  $r  \ll   \rho_4^{{5 / 2}}$.
The effective string coupling
$e^\Phi  \sim \left({\rho_4 / r}\right)^{{3/ 4}}$
is small at length scales $r  \gg \rho_4$. Thus, the type IIA supergravity description \schroiia\
is valid in the regime $ \rho_4  \ll  r  \ll \rho_4^{{7 / 3}}$. At length scales $r \ll \rho_4$ the effective string coupling is large and
we need to uplift the non-relativistic D4-brane ten-dimensional background to eleven dimensions.
In the UV we expect a five-dimensional
 non-relativistic conformal field theory description
realized on the worldvolume of the
non-relativistic M5-branes \liftdfour.

\subsec{Non-relativistic D5-branes}

The dimensionless effective coupling
$g_{eff}= {\rho_5^3 /(l_s^2r)}$ is small in the IR at length scales $r  \gg \rho_5^3$, and
we expect a perturbative  five-dimensional non-relativistic field theory description in this regime.
The description breaks down at a length scale $r  \sim \rho_5^3$.
The curvature of the non-relativistic D5-brane metric \rsce\ is small at length scales  $r  \ll \rho_5^3$.
The effective string coupling
$e^\Phi  \sim {\rho_5 / r}$
is small at length scales $r  \gg \rho_5$. Thus, the type IIB supergravity description
\solschre\ is valid in the regime $ \rho_5 \ll  r  \ll \rho_5^3$. At length scales $r \ll \rho_5$ the effective string coupling is large and
we need to apply an S-duality transformation to the non-relativistic D5-brane background.
This gives the non-relativistic NS5-branes \nsfive.
The effective string coupling now is  $e^\Phi  \sim {r / \rho_5}$ and it vanishes in the UV at small $r$.
The curvature ${\cal R} \sim {1 / \rho_5^2}$ is independent of $r$ and is small at large $\rho_5$.
Thus, the type IIB supergravity description is valid in the IR at large $\rho_5$.

\subsec{Non-relativistic D6-branes}

The dimensionless effective coupling
$g_{eff}= l_s^{-2}\left({\rho_6^7 / r^3} \right)^{1 / 2}$ is small in the IR  at length scales $r  \gg \rho_6^{{7 / 3}}$, and
we expect a perturbative  six-dimensional non-relativistic field theory description in this regime.
The description breaks down at a length scale $r  \sim  \rho_4^{{7/ 3}}$.
The curvature of the non-relativistic D6-brane metric \schroiia\ is small at length scales  $r  \ll   \rho_6^{{7 / 3}}$.
The effective string coupling
$e^\Phi  \sim \left({\rho_6 / r}\right)^{{3/ 4}}$
is small at length scales $r  \gg \rho_4$. Thus, the type IIA supergravity description \schroiia\
is valid in the regime $ \rho_4  \ll  r  \ll \rho_6^{{7 / 3}}$. At length scales $r \ll \rho_6$ the effective string coupling is large and
we need to uplift the non-relativistic D6-brane ten-dimensional background to eleven dimensions.
We get the the non-relativistic KK monopole as described in section 4.3.

\medskip\medskip
\noindent{\bf Acknowledgements}

\noindent We would like to thank S. Yankielowicz
for discussions.
This work is supported in part by the Israeli Science Foundation center of excellence,
by the Deutsch-Israelische Projektkooperation, by the US-Israel Binational Science
Foundation and
by the European Network.


\appendix{A}{Metrics on the spheres}

In this Appendix we outline the description of the odd-dimensional spheres $S^{2n+1}$ as $U(1)$ fibrations over $\PP^{n}$.

\subsec{$U(1)\hookrightarrow S^3\hookrightarrow \PP^1$}

The metric on $\PP^1$ is
\eqn\cpone{\eqalign{
ds^2_{\PP^1}=&{1\over4}\left(d\theta^2+\sin^2\theta d\phi^2\right) \ ,\cr
{\cal A}=&\half \cos\theta d\phi \ ,
}}
where $J_{\PP\,{}^1}=d{\cal A}$. The metric on $S^3$ is given by
\eqn\sthree{
ds^2_{S^3}=\left(d\chi+{\cal A}\right)^2+ds^2_{\PP\,{}^1} \ .
}

\subsec{$U(1)\hookrightarrow S^5\hookrightarrow \PP^2$}

The metric on $\PP^2$ is
\eqn\cpone{\eqalign{
ds^2_{\PP\,{}^2}=&d\mu^2+{1\over4}\sin^2\mu \left(\sigma_1^2+\sigma_2^2+\cos^2\mu\sigma_3^2\right)  \ ,\cr
{\cal A}=& \half\sin^2\mu\,\sigma_3   \ ,
}}
where $\sigma_i$ are the $SU(2)$ left invariant currents that satisfy $d\sigma_i=-\half\epsilon_{ijk}\sigma_j\wedge\sigma_k$
\eqn\leftinv{\eqalign{
\sigma_1=&\cos\psi d\theta+\sin\theta\sin\psi d\phi \ ,\cr
\sigma_2=&-\sin\psi d\theta+\sin\theta\cos\psi d\phi \ ,\cr
\sigma_3=& d\psi+\cos\theta d\phi \ ,
}}
and $J_{\PP\,{}^2}=d{\cal A}$.
The metric on $S^5$ is given by
\eqn\sthree{
ds^2_{S^5}=\left(d\chi+{\cal A}\right)^2+ds^2_{\PP\,{}^2} \ .
}

\subsec{$U(1)\hookrightarrow S^7\hookrightarrow \PP^3$}

The metric on $\PP^3$ is
\eqn\cpone{\eqalign{
ds^2_{\PP\,{}^3}=&d\mu^2+\sin^2\mu d\alpha^2+{1\over4}\sin^2\mu\Bigl[\sin^2\alpha\left(\sigma_1^2+\sigma_2^2+\cos^2\alpha\sigma_3^2\right)\cr
&+\cos^2\mu\Bigl(d\lambda^2+2\sin^2\alpha\left(d\lambda d\psi+\cos\theta d\lambda d\phi\right)\cr
&+\sin^4\alpha\left(d\psi^2+\cos^2\theta d\phi^2+2\cos\theta d\psi d\phi\right)\Bigr)\Bigr]  \ ,\cr
{\cal A}=&\half \sin^2\mu\left(d\lambda+\sin^2\alpha\sigma_3\right)   \ ,
}}
where $\sigma_i$ are the $SU(2)$ left invariant currents in \leftinv\
and $J_{\PP\,{}^3}=d{\cal A}$.
The metric on $S^7$ is given by
\eqn\sthree{
ds^2_{S^7}=\left(d\chi+{\cal A}\right)^2+ds^2_{\PP\,{}^3} \ .
}

\listrefs

\bye